\documentclass[10pt,conference]{IEEEtran}

\IEEEoverridecommandlockouts
\usepackage{amsmath,amssymb,amsfonts}
\usepackage{algorithm}
\usepackage{algorithmic}
\usepackage{graphicx, url}

\usepackage{textcomp}

\usepackage{xcolor}
\usepackage{stfloats}

\usepackage{amsmath}
\usepackage{amssymb}
\usepackage{float}
\usepackage{subcaption}
\usepackage[caption=false,font=footnotesize,labelfont=sf,textfont=sf]{subfig}  

\renewcommand{\baselinestretch}{0.922}

\usepackage[compact]{titlesec}         
\titlespacing{\section}{0pt}{0pt}{0pt} 
\AtBeginDocument{
  \setlength\abovedisplayskip{0pt}
  \setlength\belowdisplayskip{0pt}}

\ifCLASSINFOpdf
  
\else

\fi

\begin{document}

\title{Enhancing SDVN Performance via Policy-Driven Lightweight Control-Plane Resizing Strategies}
\author{{Muhammad Zain Ul Abideen, Prathapasinghe Dharmawansa, Nurul Huda Mahmood, and  
Chafika Benza\"{i}d }\\ 
     \IEEEauthorblockA{Centre for Wireless Communications, University of Oulu, Finland. \\ 
		Email: \{muhammad.zainulabideen, prathapasinghe.kaluwadevage, nurulhuda.mahmood, chafika.benzaid\}@oulu.fi. } }

\maketitle
\begin{abstract}

Software-defined vehicular networks (SDVNs) under high mobility and fluctuating traffic demand offer programmable, centralized control for latency-sensitive intelligent transportation systems. However, data-plane Quality of Service (QoS) is often degraded by control-plane overload due to frequent handovers and dense vehicle-to-infrastructure (V2I) contacts. To address this, we propose two lightweight mechanisms for low-latency control-plane resizing in multi-controller SDVNs. The first - \textit{Control-plane Centric Control-plane Resizing Mechanism} - proactively offloads roadside units from overloaded controllers to underloaded or idle ones when a predefined load threshold is exceeded, preventing prolonged overload with minimal decision latency. The second - \textit{Data-plane Centric Control-plane Resizing Mechanism} - triggers resizing based on observable data-plane QoS degradation, such as average round-trip time exceeding a QoS threshold, aligning control-plane adaptation with V2I service experience. Both mechanisms are implemented and evaluated on Mininet-WiFi emulation testbeds with realistic worst-case vehicles mobility. Compared to fixed single-controller and static multi-controller benchmarks, the proposed algorithms significantly reduce end-to-end delay and packet loss while improving load balancing rate.
\end{abstract}

\IEEEpeerreviewmaketitle

\begin{IEEEkeywords}
Control-plane load, control-plane resizing, Mininet-WiFi, quality of service, RSU offloading, SDVN
\end{IEEEkeywords}

\section{Introduction} \label{sec:intro}

Intelligent Transportation Systems (ITS) are vital for Beyond-5G and 6G ecosystems, supporting applications like cooperative driving and real-time infotainment~\cite{liu20246g}. These applications require strict Quality of Service (QoS) criteria, including low end-to-end (E2E) latency, minimal packet loss, and service continuity under high mobility. As vehicles dynamically connect to roadside units (RSU), the network must maintain reliable data-plane performance while adapting to fluctuating traffic and frequent handovers~\cite{silva2021exploring}. Software-defined vehicular networks (SDVN) can manage vehicular networks’ complexity by separating the control-plane from the data-plane~\cite{islam2021software}. In SDVN systems, RSUs provide vehicle-to-infrastructure (V2I) connectivity, while SDN controllers manage mobility, flow setup, and policy enforcement~\cite{altahrawi2021multi}. However, as vehicle density and mobility increase, the control-plane can become a performance bottleneck due to delays in processing forwarding rules, increased latency in installing flow rules at RSUs, and congestion during frequent vehicle handovers, resulting in poor data-plane QoS.

Despite the flexibility of SDVN architectures, highly dynamic vehicular environments pose challenges to control-plane scalability and QoS assurance~\cite{hussein2024sdn}. Rapid topology changes, frequent handovers, uneven vehicle density, and intermittent wireless connectivity impose heavy and time-varying processing demands on SDN controllers~\cite{he2016sdvn}, often leading to flow installation delays, packet loss, and degraded E2E performance. While multi-controller architectures improve scalability and fault tolerance, they can introduce controller load imbalance due to mobility-driven traffic skew, undermining routing efficiency and service continuity~\cite{boukerche2021design}.

Load balancing and scalability in SDNs have been addressed in different ways.  Examples include fractional switch migration for fine-grained load redistribution~\cite{prajapati2024fractionallb}, joint controller placement and routing optimization via hybrid deep reinforcement learning and graph neural networks \cite{farhan2025recap}, and programmable data-plane offloading with adaptive multi-controller coordination~\cite{alyanbaawi2025mc}. However, these approaches predominantly rely on control-plane-centric metrics, such as controller utilization, flow arrival rates, or synchronization overhead, and assume static or semi-static topologies. Such assumptions are poorly suited for SDVNs, where data-plane QoS degradation manifests itself as an increased round trip time (RTT) and packet loss~\cite{babbar2022lbsmt}. Moreover, learning-based SDVN solutions often incur high training time \cite{smida2022intelligent}. These limitations motivate the need for lightweight, mobility-aware 
control-plane resizing mechanisms that jointly consider controller load dynamics and data-plane QoS degradation in vehicular environments.

This paper introduces two lightweight, policy-based control-plane resizing mechanisms. The first, \textit{Control-plane Centric Control-plane Resizing Mechanism (CCCRM)}, proactively offloads RSUs based on controller load thresholds to balance the control-plane load. The second, \textit{Data-plane Centric Control-plane Resizing Mechanism (DCCRM)}, reactively resizes the control-plane in response to data-plane QoS degradation. \textit{CCCRM} and \textit{DCCRM} are complementary data-plane QoS-driven adaptation mechanisms that alleviate mobility-induced control-plane congestion and improve overall SDVN stability. Both mechanisms are extensively evaluated using the Mininet-WiFi emulation platform~\cite{fontes2015mininet}, demonstrating improvements in E2E delay, packet loss, and load balancing rate compared to benchmark algorithms.

The rest of the paper is organized as follows: Section \ref{sec:system Model} presents the system model design. The two proposed mechanisms are detailed in Section \ref{sec:proposed idea}. Section \ref{sec:Experiment and Results} highlights the experiment and the results of the proposed mechanisms. Section \ref{sec:conclusion} presents the conclusion of our study.

\section{System Model Design} \label{sec:system Model}

\begin{figure}[b] 
    \centering
    \includegraphics[width=0.47\textwidth, height=7.5cm]{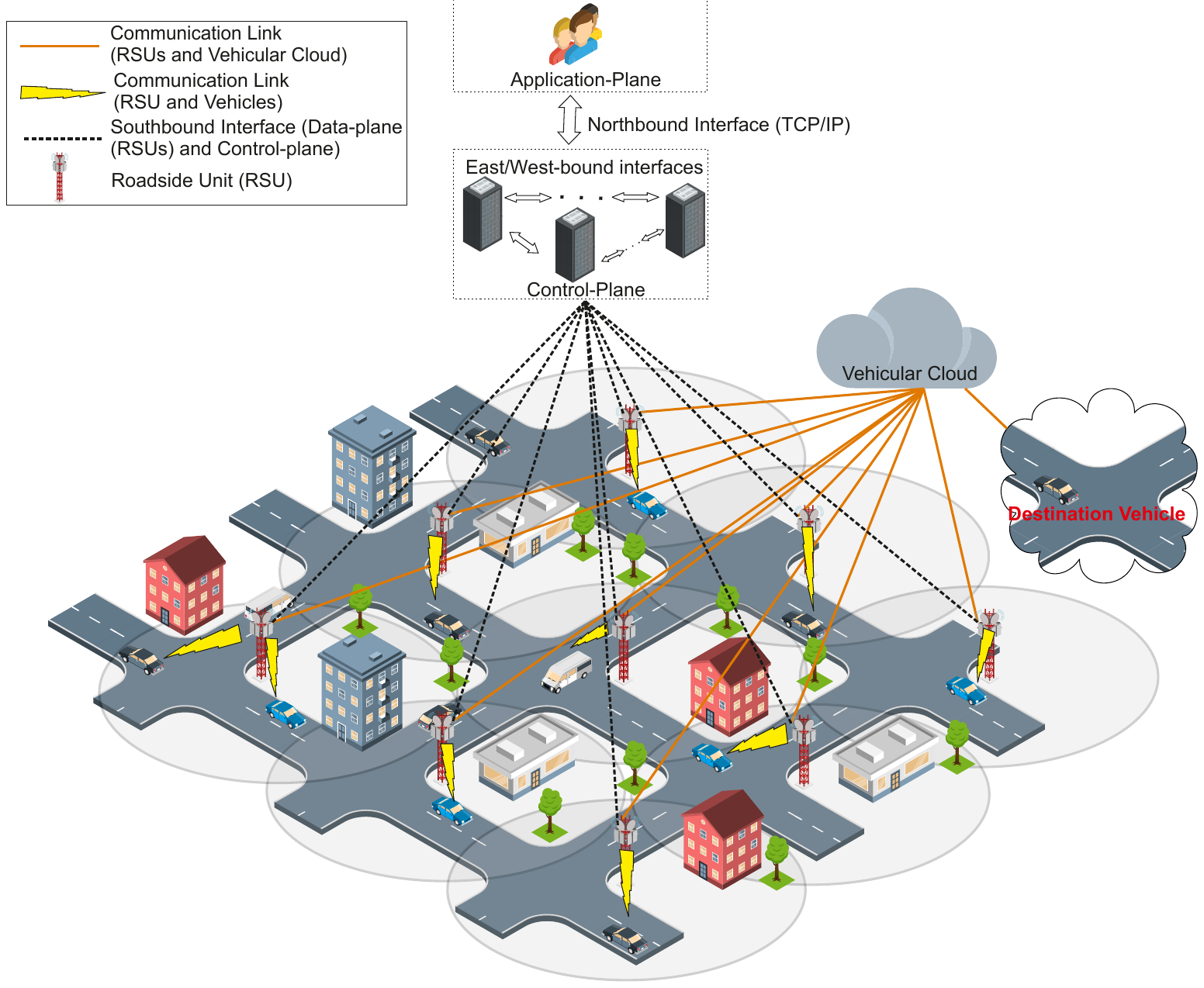}
    \caption{System model of the considered SDVN architecture.}
    \label{fig:system_model}
\end{figure}

We consider a multi-controller SDVN composed of three logical layers: the application-plane, control-plane, and data-plane. Let
$\mathcal{C} = \{C_1, C_2, \dots, C_{N_c}\}$ denote the set of SDN controllers,
$\mathcal{R} = \{R_1, R_2, \dots, R_{N_r}\}$ the set of RSUs, and
${V} = \{V_1, V_2, \dots, V_{N_v}\}$ the set of mobile vehicles, where $N_c$, $N_r$, and $N_v$ indicate the number of controllers, RSUs, and vehicles, respectively. 
The destination vehicle shown in Fig.~\ref{fig:system_model} represents a static external endpoint located outside the SDVN access network. It is connected through the wired core via a vehicular cloud and serves solely as a traffic sink for E2E data-plane performance evaluation. Mobile vehicles do not communicate with each other; however, each vehicle generates traffic only toward the fixed destination vehicle. This modeling choice isolates the impact of vehicular mobility, RSU handovers, and controller-RSU interactions on control-plane load and data-plane QoS, without introducing additional complexity from vehicle-to-vehicle communication or multi-destination traffic patterns.

\subsection{Problem Statement} \vspace{-3pt}
All RSUs may initially be assigned to a single controller (such as $C_1$). 
Vehicles move across the RSU coverage areas arbitrarily with various speeds. Each vehicle connects to the closest RSU within its communication range~\cite{abideen2023development}. 
As vehicles move, changes in signal strength trigger RSU handovers, resulting in dynamic data-plane connectivity and increased control-plane signaling. 
The dynamic imbalance of the controller load, which results from both static RSU-controller mappings and time-varying RSU-vehicle associations, is the main issue this work attempts to address. Our goal is to determine which, when and how many RSU(s) should be offloaded and redistributed among the controllers to minimize control-plane overload, satisfy data-plane QoS constraints, and prevent unnecessary controller reconfiguration and synchronization overhead, thereby reducing signaling load. Moreover, the proposed RSUs offloading decisions must be designed to be computationally efficient and lightweight to support real-time operation.  

\vspace{2mm}

\section{Proposed Control-plane resizing Algorithms} 
\label{sec:proposed idea}

\begin{figure}[b]
  \centering
  \includegraphics[width=0.75\linewidth]{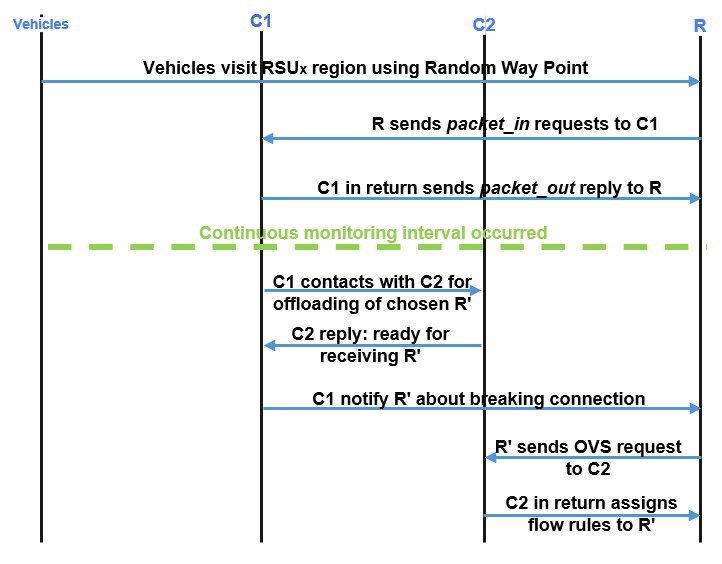}
  \caption{General timing diagram of control-plane resizing.}
  \label{TimingDiagram}
\end{figure}

The proposed control-plane resizing algorithms adapt to vehicular networks' dynamic conditions by using observable metrics like controller load and QoS degradation for real-time decisions. Threshold-based triggers, periodic checks, and incremental RSU offloading ensure stability, avoiding oscillations. Within a closed-loop framework, the RSU-centric process redistributes control tasks locally without relocating controllers or reconfiguring the topology globally. 
To highlight the general RSUs reassignment and control-plane resizing process, Fig.~\ref{TimingDiagram} provides a general timing diagram elaborating the sequence of interactions between vehicles, RSUs, and multiple controllers. 
RSUs $\mathcal{R}$ forward packets according to set flow rules, requesting controller intervention only when necessary. Every controller locally assesses its control-plane load and related data-plane performance parameters for the RSUs under its control at periodic monitoring intervals. The affected controller is identified differently in \textit{CCCRM} and \textit{DCCRM}. In \textit{CCCRM}, any controller whose measured load exceeds a predefined threshold is considered affected. In \textit{DCCRM}, the affected controller corresponds to the bottleneck controller, defined as the controller with the highest average RTT that violates the specified threshold. Once identified, the affected controller automatically starts a resizing operation by cooperating with other underloaded controllers via east-west communication when overload or QoS violations are detected. The proposed algorithm balances the controller load by reassigning selected RSUs $\mathcal{R}'$ $\subset$ $\mathcal{R}$ under the overloaded controller to new (currently underloaded) controllers. During the controller-RSU reassignment process, ongoing V2I traffic experiences a temporary disruption, as a hard handover regime is assumed. 
Packets generated during this process are buffered at the vehicle's application layer instead of being instantly dropped once the new controller-RSU connection is established \cite{abideen2023development}. Both proposed resizing algorithms are based on this general abstraction, with their primary differences lying in how the overload conditions are determined. 

\subsection{Control-plane Centric Control-plane Resizing Mechanism} \label{sec:CCCRM algo}

In this algorithm, the controller load is collected through periodic sampling. Incoming packets from an RSU without a corresponding forwarding rule are known as $packet\_in$ messages. Their handling constitutes the dominant source of computational load on the controller. To quantify the controller's load, let $L_{c,k}$ denote the load of controller $c$ observed during the monitoring interval $k$, corresponding to a fixed time window $t$, which can be 
mathematically expressed as
\begin{equation}
L_{c,k} = \sum_{r \in \mathcal{R}_{c,k}} pi_{r,k} \, ,
\label{eq:controller_load}
\end{equation}
where $pi_{r,k}$ represents the number of $packet\_in$ messages generated by RSU $r$ during interval $k$. 
A controller is considered overloaded when its load exceeds a predefined threshold $L_{\text{th}}$; i.e., $L_{c,k} > L_{\text{th}}$. In practical applications, $L_{\text{th}}$ corresponds to the maximum viable control-plane load based on controller resource constraints (e.g., central processing unit usage or flow rules processing rate) and can be determined using offline benchmarking or capacity profiling to ensure reliable performance.
Once an overload condition is detected, the average control-plane load contributed by a single RSU in the interval $k$ is approximated as 
\begin{equation}
\bar{L}_{r,c,k} \approx \frac{L_{c,k}}{|\mathcal{R}_{c,k}|} \, ,
\label{eq:avg_load_rsu}
\end{equation}
where $|\mathcal{R}_{c,k}|$ denotes the maximum number of RSUs connected to the controller $c$ during the interval $k$.
The severity of controller overload at a given monitoring interval $k$ is quantified as
\vspace{-0.18cm}
\begin{equation}
\Delta L_{c,k} = L_{c,k} - L_{\text{th}}.
\label{eq:excess_load}
\end{equation}
This allows \textit{CCCRM} to scale its response proportionally to the overload intensity instead of triggering an abrupt offloading action.
After initiating an offloading decision, the minimum number of RSUs to be offloaded from overloaded controller $c$ at interval $k$ is determined as
\begin{equation}
N^{\text{off}}_{c,k}
=
\left\lceil
\alpha \cdot
\frac{\Delta L_{c,k}}{\bar{L}_{r,c,k}}
\right\rceil,
\label{eq:num_offloaded_rsus}
\end{equation}
where $\alpha \in (0,1]$ is a tunable scaling parameter that controls the aggressiveness of the offloading process and $\lceil \cdot \rceil$ is the ceiling function enforcing an integer-valued decision. 

The RSUs to be offloaded are then selected based on vehicular mobility characteristics rather than load ranking, as all RSUs connected to the same controller are assumed to contribute equally to the control-plane load, which is approximated by the average per-RSU load given by~\eqref{eq:avg_load_rsu}. Thus, RSU selection is performed based on minimum session time. This metric is defined as the duration for which a vehicle remains within the communication coverage of an RSU after initial entry~\cite{abideen2024evaluation}. It is initially expressed as $\mathcal{N}(r) = \{ r' \in \mathcal{R} \mid r' \text{ is a neighboring RSU of } r \}$. The average session duration of vehicles that transition from a neighboring RSU $r'$ to RSU $r$ during the monitoring interval $k$ is denoted as $T_{r',r,k}$. For each RSU $r$, the minimum average session duration with its neighboring RSU is defined as
\begin{equation}
\underline{T}_{r,k}
=
\min_{r' \in \mathcal{N}(r)} T_{r',r,k},
\label{eq:min_session_time}
\end{equation}
where $\underline{T}_{r,k}$ represents the minimum average session duration observed at RSU $r$ during the monitoring interval $k$, considering all its neighboring RSUs. By taking the minimum value over all neighboring RSUs, $\underline{T}_{r,k}$ captures the fastest mobility flow associated with RSU $r$. The subsequent step involves sorting the RSUs in descending order based on their calculated average session time. This ranking prioritizes RSUs with higher minimum session time durations, as they indicate lower relative vehicular mobility and less handovers, making them suitable candidates for offloading. Thus, the updated sorted RSUs list for offloading is updated in ${R}^{\text{off}}_{c,k}$. 
The total control-plane load removed through offloading is given by
\begin{equation}
L^{\text{off}}_{c,k}
=
N^{\text{off}}_{c,k}
\cdot
\bar{L}_{r,c,k}.
\label{eq:offloaded_load}
\end{equation}
This value represents the expected reduction in controller load if all selected RSUs are successfully migrated. The number of underloaded controllers required to accommodate the offloaded RSUs is determined by checking whether the total load to be offloaded satisfies the $L^{\text{off}}_{c,k} \ge \Delta L_{c,k}$ condition. The RSUs selected for offloading are reassigned to available underloaded controllers via east-west communication, ensuring that each RSU is associated with exactly one controller. The complete procedure is summarized in Algorithm~\ref{alg:CCCRM}. 

\textbf{Complexity Analysis:} The dominant operation in Algorithm~\ref{alg:CCCRM} is sorting RSUs associated with the overloaded controllers based on session time at each interval $k$. In the worst case, this involves sorting all RSUs, resulting in a complexity of $\mathcal{O}(k \, C \, + k \, C_o \, OSort)$ where $OSort$ follows the merge sort performed on RSUs, which is generally represented as $\mathcal{O}(n\log n)$. Finally, the effectiveness of \textit{CCCRM} is evaluated using the Load Balancing Rate ($\mathrm{LBR}$), which measures the uniformity of controller load distribution over time. A higher $\mathrm{LBR}$ value indicates a more uniform control-plane load (i.e., the RTT-induced load) distribution since overloaded controllers increase data-plane RTT delay through delayed flow-rule installation and frequent handovers. On the other hand, a value of $\mathrm{LBR}$ close to zero reflects the absence of load balancing capability. To provide an overall assessment of load balancing performance, the instantaneous controller load distribution is used to calculate the $\mathrm{LBR}$ at each monitoring interval, which are then averaged over the entire duration. The $\mathrm{LBR}$ of the control-plane after offloading RSUs from an overloaded controller to underloaded ones is computed as:
\begin{equation}
\mathrm{LBR} = \frac{1}{T} \sum_{t=1}^{T} 
\left(
1 - 
\frac{\sum_{i=1}^{N_c} \left| L(c_i,t) - \bar{L}(t) \right|}
{N_c \cdot \bar{L}(t)}
\right),
\end{equation}
where $L(c_i,t)$ represents the load of each controller at time $t$, $\bar{L}(t)$ denotes the average load of controllers at time instant $t$, and $T$ is the total simulation time. 

\begin{algorithm}[t]
\caption{CCCRM}
\label{alg:CCCRM}
\textbf{Input:}
Controller set $\mathcal{C}$; 
RSU set $\mathcal{R}$; 
controller load threshold $L_{\text{th}}$; 
monitoring interval $k$; 
scaling factor $\alpha$;\\
\textbf{Output:}
Balanced controller load and updated RSU-to-controller mapping;
\begin{algorithmic}[1]
\STATE Initialization: Overloaded controller set $\mathcal{C}_o = \emptyset$, Underloaded controller set $\mathcal{C}_u = \emptyset$;\\
\FOR{each monitoring interval $k$}
    \FOR{each controller $c \in \mathcal{C}$}
        \STATE Measure controller load $L_{c,k}$ using Eq.~\eqref{eq:controller_load};
        \IF{$L_{c,k} > L_{\text{th}}$}
            \STATE $\mathcal{C}_o \leftarrow \mathcal{C}_o \cup \{c\}$;
        \ELSE
            \STATE $\mathcal{C}_u \leftarrow \mathcal{C}_u \cup \{c\}$;
        \ENDIF
    \ENDFOR

    \FOR{$c \in \mathcal{C}_o$}
        \STATE Compute $\bar{L}_{r,c,k}$ using Eq.~ \eqref{eq:avg_load_rsu};
        \STATE Compute $\Delta L_{c,k}$ using Eq. \eqref{eq:excess_load};
        \STATE Compute $N^{\text{off}}_{c,k}$ using Eq.~ \eqref{eq:num_offloaded_rsus};
        \STATE Rank RSUs connected to $c$ by minimum session time using Eq.~\eqref{eq:min_session_time} and sort them in descending order;
        \STATE Select $N^{\text{off}}_{c,k}$ RSUs for offloading $\mathcal{R}^{\text{off}}_{c,k}$ RSUs;
        \STATE Compute $L^{\text{off}}_{c,k}$ using Eq.~\eqref{eq:offloaded_load};
        \IF{$L^{\text{off}}_{c,k} \le \Delta L_{c,k}$}
            \STATE Select a single target controller $c' \in \mathcal{C}_u$;
            \STATE Offload RSUs in $\mathcal{R}^{\text{off}}_{c,k}$ to $c'$;
            \STATE Update controller-RSU assignment;
        \ELSE
            \STATE Select multiple underloaded controllers $\mathcal{C}'_u \subseteq \mathcal{C}_u$;
            \STATE Offload RSUs in $\mathcal{R}^{\text{off}}_{c,k}$ to $\mathcal{C}'_u$;
            \STATE Update controller-RSU assignment;
        \ENDIF
    \ENDFOR
\ENDFOR
\RETURN Updated controller load distribution and RSU assignments;
\end{algorithmic}
\end{algorithm}
\subsection{Data-plane Centric Control-plane Resizing Mechanism} \label{sec:DCCRM Algo}


\textit{DCCRM} addresses cases where control-plane metrics appear stable but data-plane QoS (e.g., V2I delay) degrades in dynamic vehicular networks. Unlike \textit{CCCRM}, which relies solely on control-plane load, \textit{DCCRM} uses real-time RTT and packet loss measurements from vehicles to trigger resizing when QoS thresholds are violated. In this approach, each controller periodically computes the average RTT across its RSUs, identifies the bottleneck controller with the highest RTT, and offloads the most stressed RSUs to underloaded controllers using a score-based assignment when QoS thresholds are exceeded. The average round-trip delay encountered by vehicles connected to RSU $r$ during the monitoring interval $k$ is represented by the RSU-level RTT $\bar{rtt}_{r,k}$. It is calculated by averaging the RTT samples reported by all vehicles $v \in V_{r,k}$ associated with RSU $r$ as 
\begin{equation}
\bar{rtt}_{r,k} = \frac{1}{|V_{r,k}|} \sum_{v \in V_{r,k}} RTT_v, \quad \forall r \in \mathcal{R}.
\label{eq:DCCRM_RSURTT}
\end{equation}
Thereafter, controller-level QoS is obtained by aggregating its corresponding RSU-level RTTs 
\begin{equation}
\bar{rtt}_{c,k} = \frac{1}{|\mathcal{R}_{c,k}|} \sum_{r \in \mathcal{R}_{c,k}} \bar{rtt}_{r,k},  \quad \forall c \in \mathcal{C}.
\label{eq:DCCRM_ControllerRTT}
\end{equation}

The synchronization cost of the controller $c$ at time $t_k$ is denoted by $s_c(t_k)$. This term captures the control-plane overhead resulting from RSU reassignment, including inter-controller coordination and flow rule reinstallation, and represents the temporary delay introduced by resizing operations. It is used to reduce or minimize the occurrence of excessive resizing operations. The synchronization cost of controller $c$ in the previous interval $t_{k-1}$ is denoted by $s_c(t_{k-1})$. The incremental synchronization overhead caused by controller-RSU reassignment and the controller $c$ experiences between two consecutive monitoring intervals is represented as $\Delta s_{c,k}$ which is computed as
\begin{equation}
\Delta s_{c,k} = s_c(t_k) - s_c(t_{k-1}), \quad \forall c \in \mathcal{C}.
\label{eq:DCCRM_Cont.SyncDelay}
\end{equation}
The bottleneck controller $c^*$ at monitoring interval $k$ is identified as the controller experiencing the highest average RTT. The bottleneck controller $c^*$ is identified as
\begin{equation}
c^* =
\arg\max_{c \in \mathcal{C}}
\bar{rtt}_{c,k}.
\label{eq:DCCRM_BottleneckController}
\end{equation}
Control-plane resizing is triggered only when the bottleneck controller violates the predefined RTT threshold, and resizing is triggered only if $\bar{rtt}_{c^*,k} > rtt_{\mathrm{th}}$.

The offload size $n_k$ scales with both the normalized RTT excess and the number of RSUs currently managed by the bottleneck controller, ensuring adaptive and incremental resizing. Thus, the number of RSUs to be offloaded at the monitoring interval $k$ is computed as
\begin{equation}
n_k =
\left\lceil
\kappa \cdot 
\frac{\bar{rtt}_{c^*,k} - rtt_{\mathrm{th}}}{rtt_{\mathrm{th}}}
\cdot
|\mathcal{R}_{c^*,k}|
\right\rceil.
\label{eq:DCCRM_No.RSUsOffload}
\end{equation}
The parameter $\kappa$ regulates the aggressiveness of the resizing process.
RSUs connected to the bottleneck controller are assigned a weight using a stress metric that captures both delay and traffic density, i.e., 
\begin{equation}
P_{r,k} = \bar{rtt}_{r,k} \cdot |V_{r,k}|.
\label{eq:DCCRM_PressureMetric}
\end{equation}
The top-$n_k$ RSUs with the highest stress connected to $c^*$ are then selected
\vspace{-0.18cm}
\begin{equation}
\mathcal{R}^{off}_k =
\text{top-}n_k
\left(
\arg\max_{r \in \mathcal{R}_{c^*,k}} P_{r,k}
\right).
\label{eq:DCCRM_OffloadRSUs}
\end{equation}
After identifying the bottleneck controller and the selected RSUs to offload, these RSUs are offloaded to underloaded controllers based on a variable $Score_{c,k}$ representing their load, which is calculated for each candidate target controller $c \neq c^*$ as follows
\vspace{-0.18cm}
\begin{equation}
Score_{c,k} =
w_L |\mathcal{R}_{c,k}|
+
w_S \frac{\Delta s_{c,k}}{S_{\mathrm{th}}},
\label{eq:DCCRM_ScoreUnderloadController}
\end{equation}
where the weighting factors $w_L$ and $w_S$ balance controller load and  synchronization cost. $S_{\mathrm{th}}$ is the maximum permissible synchronization cost used for normalization. 
The target controllers are ranked in ascending order of their current load and updated as the sorted set ${C}^{sorted}_k$. The selected RSUs are assigned to candidate underloaded controllers using a modulo-based controlled round-robin strategy $\phi_k(r_i)$ over the ranked controller sorted in ascending order based on $Score_{c,k}$. This ensures a fair distribution of offloaded RSUs while preventing secondary controller overload. After migration, the controller-RSU mappings are updated accordingly. The pseudocode of \textit{DCCRM} is presented in Algorithm \ref{alg:DCCRM}. 

\textbf{Complexity Analysis:} The dominant operations in Algorithm~\ref{alg:DCCRM} include computing RTT metrics across RSUs, sorting RSUs associated with the bottleneck controller for offloading, and sorting underloaded candidate controllers at each interval $k$. This yields a complexity of $\mathcal{O}(k \, R + k \, C + k \, OSort)$, where $OSort$ follows a merge sort applied to RSUs associated with overloaded controllers, among underloaded controllers, and assignment of chosen RSUs to one or more underloaded controllers, generally represented as $\mathcal{O}(n\log n)$. 

The effectiveness of \textit{DCCRM} is evaluated using the metric data-plane $\mathrm{(DP)\text{-}LBR}$, which captures the degree of data-plane load balance among active controllers by measuring the normalized deviation of controller loads from their mean value, and is defined as
\begin{equation}
\mathrm{DP\text{-}LBR}(k) =
\begin{cases}
0 & \text{if } N_c(k) = 1, \\
1 - \dfrac{\sum\limits_{c \in \mathcal{C}_k} \left| L(c,k) - \bar{L}(k) \right|}{N_c(k) \cdot \bar{L}(k)} & \text{if } N_c(k) \ge 2,
\end{cases}
\label{eq:DP-LBR}
\end{equation}
where $\mathcal{C}_k$ is the set of active controllers, $N_c(k) = |\mathcal{C}_k|$ is the number of active controllers who have RSU(s) under their domain, $L(c,k)$ is the load of controller $c$ in interval $k$ (e.g., sum of RTTs $\bar{rtt}_{c,k} $), and $\bar{L}(k)$ is the average load (from the viewpoint of RTT) across active controllers given by
    \begin{equation}
        \bar{L}(k) = \frac{1}{N_c(k)} \sum_{c \in \mathcal{C}_k} \bar{rtt}_{c,k}. \nonumber
    \end{equation}
Given the long duration of traffic observation and dynamic changes in traffic and controller count, the time-averaged $\mathrm{DP\text{-}LBR}$ ($\mathrm{DP\text{-}LBR}_{\mathrm{avg}}$) is adopted as a single scalar metric to assess the overall effectiveness and consistency of data-plane load balancing throughout the simulation.
The overall load balancing performance across all intervals is computed by averaging over intervals with at least two active controllers
\begin{equation}
\mathrm{DP\text{-}LBR}_{\mathrm{avg}} =
\frac{1}{|\mathcal{T}^*|} \sum_{k \in \mathcal{T}^*} \mathrm{DP\text{-}LBR}(k),
\end{equation}
where $\mathcal{T}^* = \{ k \mid N_c(k) \ge 2 \}.$ 
This formulation ensures that intervals with only a single controller, where balancing is not defined, are excluded.
\vspace{2mm}
\section{Experiment setting and results discussion} \label{sec:Experiment and Results}
In the conducted experimentation, the topology design illustrated in Fig.~\ref{fig:system_model} is implemented using a virtual machine (VM) of Ubuntu 20.04.02 operating system on 11th Gen Intel Core i5 CPU with 4GB of RAM. The simulation of the SDN based vehicular network is performed using Mininet-WiFi 2.6 and the Python-based Ryu controller, which supports the OpenFlow v1.3 protocol. Python version considered for this work is 3.8.10. Each testbed simulation takes about one hour and the monitoring interval $k$ occurred about every $300$ seconds (s). 
A small-scale topology with $N_c = 3$, $N_r = 9$, and $N_v = 10$ is designed to evaluate the two proposed lightweight control-plane resizing mechanisms. This topology size chosen intentionally within the practical resource limits of Mininet-WiFi emulation running on a single VM. Because both \textit{CCCRM} and \textit{DCCRM} are threshold-based and generalized, the fundamental behaviour and performance trends observed in this study remain consistent even for large-scale network because the algorithms' core decision logic is independent of the total number of controllers, RSUs, or vehicles. Each vehicle generates a packet at each $0.2 s$ interval. The considered mobility model for vehicles is random way point model~\cite{lin2013_RWP} and the propagation model adopted is Friis propagation loss as provided by Mininet-WiFi. 

\begin{algorithm}[t]
\caption{DCCRM}
\label{alg:DCCRM}
\textbf{Input:}
Controller set $\mathcal{C}$;
RSU set $\mathcal{R}$;
RTT threshold $rtt_{\mathrm{th}}$;
monitoring interval $k$;
resizing factor $\kappa$;
weights $w_L, w_S$

\textbf{Output:}
QoS-aware balanced controller--RSU mapping;
\begin{algorithmic}[1]
\STATE Initialization: Overloaded controller set $\mathcal{C}_o = \emptyset$, Underloaded controller set $\mathcal{C}_u = \emptyset$;

\FOR{each monitoring interval $k$}

    \FOR{each RSU $r \in \mathcal{R}$}
        \STATE Compute $\bar{rtt}_{r,k}$ using Eq. \eqref{eq:DCCRM_RSURTT};
    \ENDFOR

    \FOR{each controller $c \in \mathcal{C}$}
        \STATE Compute $\bar{rtt}_{c,k}$ using Eq. \eqref{eq:DCCRM_ControllerRTT};
        \STATE Compute $\Delta s_{c,k}$ using Eq. \eqref{eq:DCCRM_Cont.SyncDelay};
    \ENDFOR
    \STATE Identify bottleneck controller $c^*$ using Eq. \eqref{eq:DCCRM_BottleneckController};
    \IF{$\bar{rtt}_{c^*,k} \le rtt_{\mathrm{th}}$}
        \STATE QoS satisfied; retain current controller--RSU assignment;
    \ELSE
        \STATE $\mathcal{C}_o \leftarrow \{ c^* \}$;
        \STATE $\mathcal{C}_u \leftarrow \mathcal{C} \setminus \{ c^* \}$;

        \STATE Compute $n_k$ to offload RSUs using Eq. \eqref{eq:DCCRM_No.RSUsOffload};

        \FOR{each RSU $r \in \mathcal{R}_{c^*,k}$}
            \STATE Compute $P_{r,k}$ using Eq. \eqref{eq:DCCRM_PressureMetric};
        \ENDFOR

        \STATE Select $\mathcal{R}^{off}_k$ for offloading using Eq. \eqref{eq:DCCRM_OffloadRSUs};
        \FOR{each controller $c \in \mathcal{C}_u$}
            \STATE Compute $Score_{c,k}$ using Eq. \eqref{eq:DCCRM_ScoreUnderloadController};
        \ENDFOR
        \STATE Sort controllers in $\mathcal{C}_u$ in ascending order $\mathcal{C}^{sorted}_k$;

        \STATE Initiate assignment operation among $\mathcal{C}_u$;
        \FOR{$r_i \in \mathcal{R}^{off}_k$}
            \STATE Start assignment $\phi_k(r_i)$;
        \ENDFOR

        \STATE Update $c^*$ mappings;
        \STATE Update mappings of controllers in $\mathcal{C}_u$;
    \ENDIF
\ENDFOR
\RETURN Updated controller--RSU assignment with restored data-plane QoS;

\end{algorithmic}
\end{algorithm}

We compare the proposed mechanisms against two baselines: a single-controller testbed (all nine RSUs statically assigned to one SDN controller) and a static multi-controller design (nine RSUs uniformly divided among three controllers, never reassigned). These extremes represent fully centralized control and traditional static load distribution.

\begin{figure}[b]
\centering
\begin{subfigure}{0.49\linewidth}
    \centering
    \includegraphics[width=\linewidth]{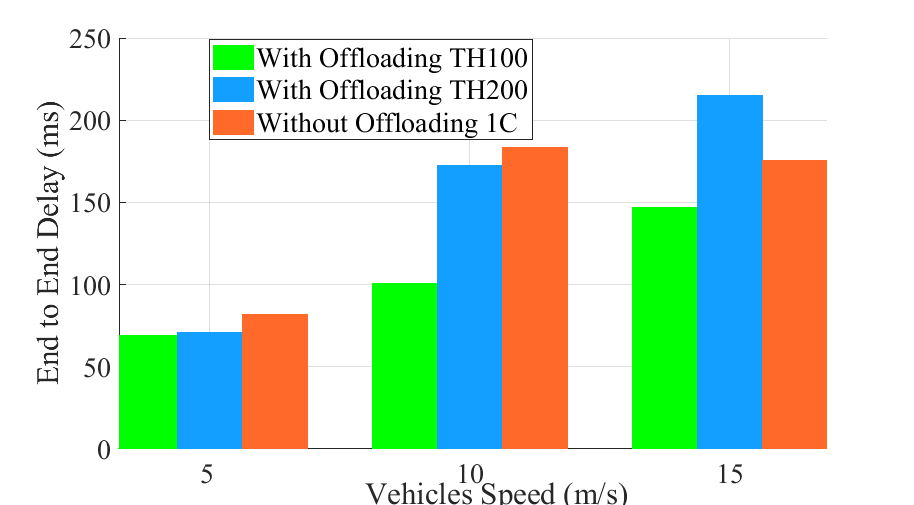}
    \caption{}
    \label{fig:CCCRM_E2EDelay}
\end{subfigure}
\hfill
\begin{subfigure}{0.49\linewidth}
    \centering
    \includegraphics[width=\linewidth]{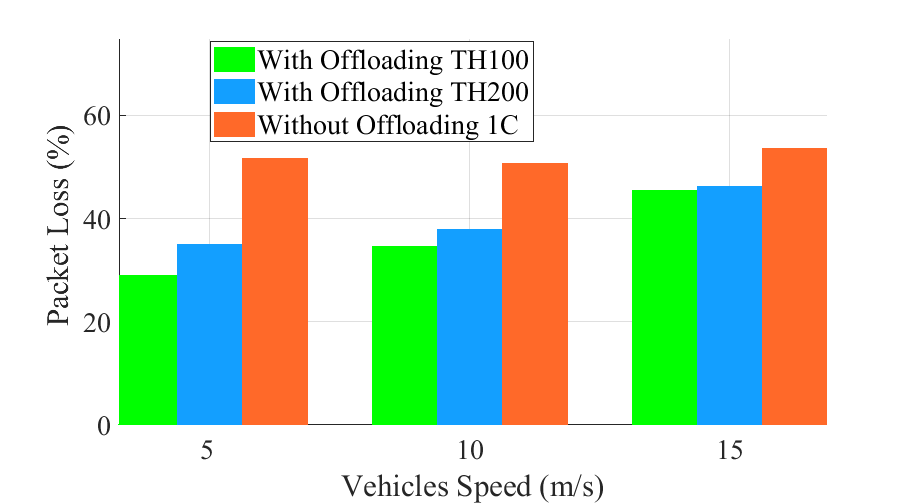}
    \caption{}
    \label{fig:CCCRM_PacketLoss}
\end{subfigure}
\caption{CCCRM: (\ref{fig:CCCRM_E2EDelay}) E2E delay with variable vehicles speed; (\ref{fig:CCCRM_PacketLoss}) Packets loss with variable vehicles speed.}
\vspace{-0.7cm}
\end{figure}

\begin{figure}[t]
\centering
\begin{subfigure}{0.49\linewidth}
    \centering
    \includegraphics[width=\linewidth]{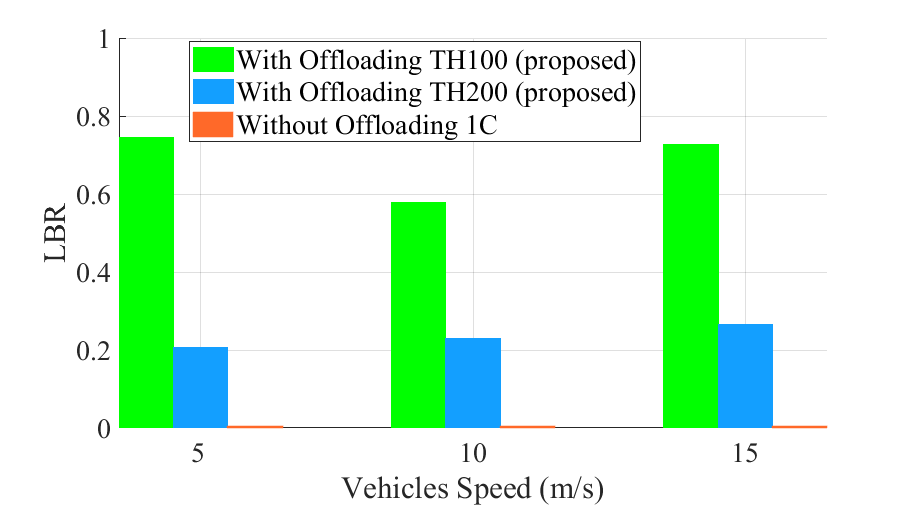}
    \caption{}
    \label{fig:lbr_CCCRM}
\end{subfigure}
\hfill
\begin{subfigure}{0.49\linewidth}
    \centering
    \includegraphics[width=\linewidth]{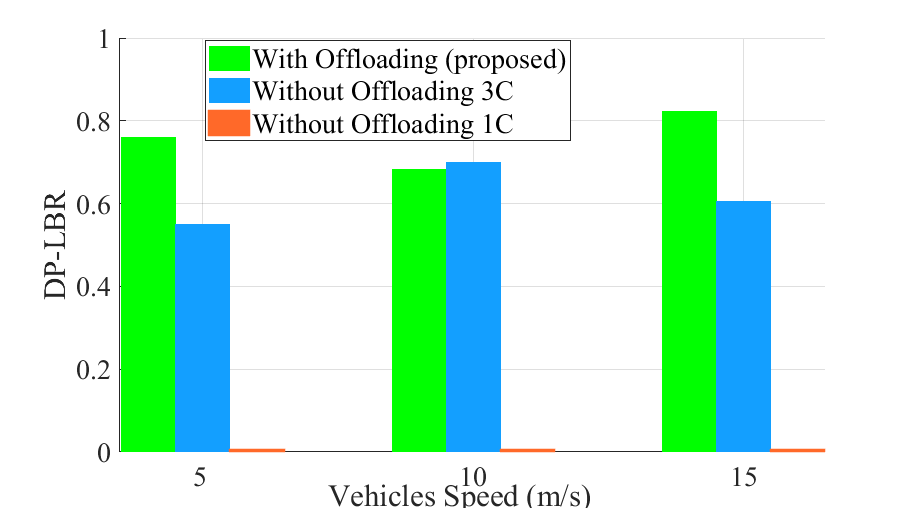}
    \caption{}
    \label{fig:lbr_DCCRM}
\end{subfigure}
\caption{LBR (\ref{fig:lbr_CCCRM}) CCCRM-LBR; (\ref{fig:lbr_DCCRM}) DCCRM-LBR }
\label{fig:LBR}
\end{figure}

As shown in Fig.~\ref{fig:CCCRM_E2EDelay}, Fig.~\ref{fig:CCCRM_PacketLoss}, and Fig.~\ref{fig:lbr_CCCRM}, \textit{CCCRM} significantly improves performance over the single-controller benchmark in most cases. In the benchmark, all RSUs remain statically assigned to one controller, causing severe control-plane overload under high mobility. This results in E2E delay increasing from 81.7 ms at 5 m/s to over 183 ms at 10 m/s and packet loss exceeding 50\%. With the aggressive threshold ($L_{\mathrm{th}}=100$), \textit{CCCRM} consistently achieves the best results, reducing E2E delay to 69.0 ms, 100.4 ms, and 147.2 ms, and packet loss to 29.0\%, 34.5\%, and 45.5\% at 5, 10, and 15 m/s, respectively. The more cautious threshold ($L_{\mathrm{th}}=200$) also outperforms the benchmark at lower speeds (5 and 10 m/s), but at the highest speed of 15 m/s, it shows a slightly higher E2E delay than the benchmark (approximately 215 ms versus 178 ms). This occurs because the higher threshold delays offloading, allowing temporary control-plane overhead to accumulate. When migration finally occurs, it introduces additional synchronization overhead and flow-rule reinstallation delays.

Despite this, \textit{CCCRM} achieve substantially higher LBR above 0.57 for $L_{\mathrm{th}}=100$, compared to nearly zero in the benchmark. 
The results show that \textit{CCCRM} effectively reduces control-plane congestion and stabilizes data-plane QoS under higher mobility, with aggressive thresholds yielding the most consistent gains. However, excessively low thresholds may cause frequent RSU reassignments, increasing synchronization and migration overhead including flow rule reinstallation and controller coordination delays, which can introduce temporary latency and instability. The threshold values chosen in this work shows a balanced operation.

\begin{figure}[t]
\centering
\begin{subfigure}{0.49\linewidth}
    \centering
    \includegraphics[width=\linewidth]{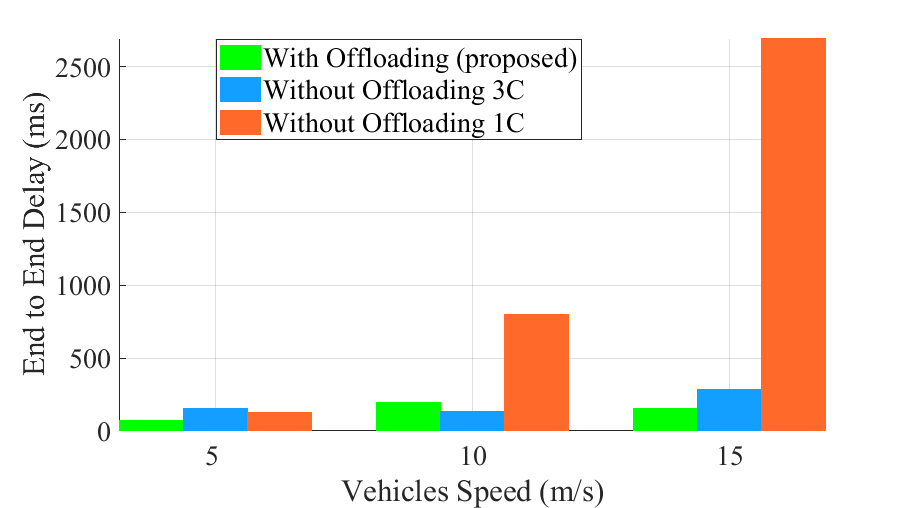}
    \caption{}
    \label{fig:DCCRM_E2EDelay}
\end{subfigure}
\hfill
\begin{subfigure}{0.49\linewidth}
    \centering
    \includegraphics[width=\linewidth]{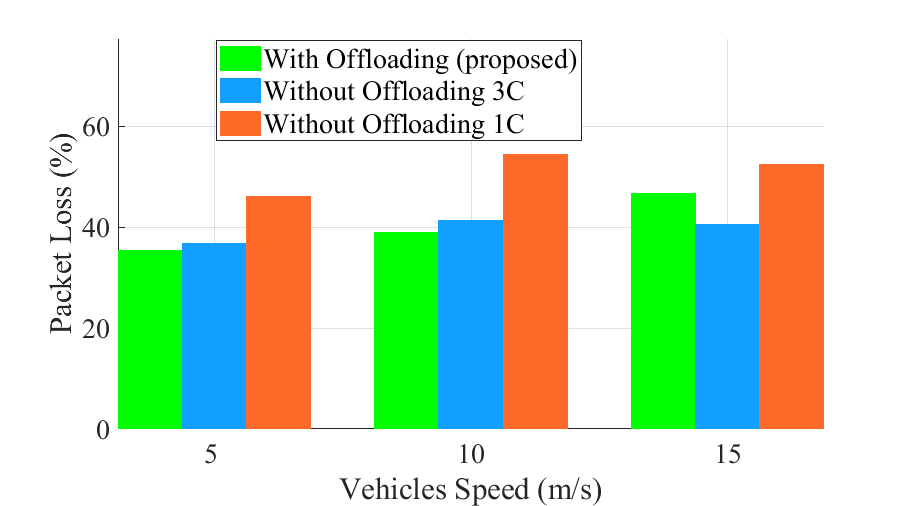}
    \caption{}
    \label{fig:DCCRM_PacketLoss}
\end{subfigure}
\caption{DCCRM: (\ref{fig:DCCRM_E2EDelay}) E2E delay with variable vehicles speed; (\ref{fig:DCCRM_PacketLoss}) Packets loss with variable vehicles speed.}
\label{fig:DCCRM_RTT_PacketsLoss}
\vspace{-0.2cm}
\end{figure}

As shown in Fig.~\ref{fig:DCCRM_E2EDelay}, Fig.~\ref{fig:DCCRM_PacketLoss}, and Fig.~\ref{fig:lbr_DCCRM}, \textit{DCCRM} improves data-plane QoS by making resizing decisions based on real-time RTT degradation rather than static controller assignments. In the single-controller benchmark, mobility-induced flow setups accumulate at one controller, causing excessive \(packet\_in\) processing and severe RTT escalation (up to $>$ 2.5 s at 15 m/s). Although the static three-controller (3C) benchmark initially distributes RSUs evenly, it cannot adapt to dynamic traffic asymmetry caused by mobility. In comparison to both static benchmarks, \textit{DCCRM} achieves the lowest E2E delay (70.1 ms) and packet loss at 5 m/s. At 10 m/s, it exhibits a somewhat higher delay than the static 3C benchmark. Because during RSU reassignment, synchronization and migration overhead introduced short-term latency before the system stabilizes, which results in a temporary increase in RTT. At the highest speed of 15 m/s, \textit{DCCRM} maintains a reasonable delay of approximately 155 ms while the single-controller scenario collapses to $>$ 2.5 s. The \(\mathrm{DP\text{-}LBR}_{\mathrm{avg}}\) results further confirm these gains, reaching up to 0.82 with the proposed mechanism compared to 0.61 for the static 3C and nearly zero for the single-controller case. These findings indicate that QoS-driven resizing offers stability and sustained performance under extremely dynamic SDVN conditions, while static load balancing based topology may overestimate system robustness. This study adopts fixed threshold values to initiate control-plane resizing and offloading decisions. Future research work will investigate adaptive threshold selection mechanisms at each occurred monitoring interval to dynamically adjust threshold values according to varying traffic loads, controller utilization, and network conditions. 

\section{Conclusion} \label{sec:conclusion}
This paper addresses controller load imbalance and overload in SDVNs under dynamic mobility. 
Two lightweight, policy-driven control-plane resizing strategies are proposed. \textit{CCCRM} uses control-plane load indicators to accomplish threshold-based reactive offloading, whereas \textit{DCCRM} uses real-time RTT degradation to initiate data-plane-aware resizing. Both algorithms are suitable for real-time execution and scale efficiently with network size. They have been extensively evaluated using Mininet-WiFi. Experimental results show that even in multi-controller deployments, static controller assignment fails to effectively handle mobility-induced traffic asymmetry. 
In contrast, the proposed strategies provide higher LBR while significantly reducing E2E delay and packet loss when compared to both static single- and multi-controller benchmarks. 
The current evaluation indicates the efficacy of the proposed mechanisms within the specified considered topology. However, larger and denser SDVN deployments may incur increased control-plane coordination overhead cost, especially under static benchmark configurations. Future research will examine such scalability and overhead-related aspects in large-scale topology scenarios and with realistic controlled vehicular mobility model. 

\section*{Acknowledgment}

{\small This work was supported by the Research Council of Finland through the projects 6G Flagship (369116), ReWin-6G (357120) and 6G-Concorse (359850).}

\bibliographystyle{IEEEtran}
\bibliography{main_final}
\end{document}